\documentclass[12pt]{iopart}

\usepackage[pdftex]{graphicx}

\usepackage{iopams}  

\usepackage{amssymb}
\usepackage{amsfonts}
\usepackage{multicol}
\usepackage{braket}

\usepackage{ascmac}
\usepackage{fancybox}
\usepackage{comment}

\usepackage{cite}

\usepackage{setstack}

\newcommand{\eq}[1]{\begin{equation} #1 \end{equation}}
\newcommand{\eqa}[2]{\begin{equation} #1 \label{#2} \end{equation}}
\newcommand{\balign}[1]{\begin{eqnarray} #1 \end{eqnarray}}

\newcommand{\fn}{\footnote}

\newcommand{\figin}[4]
{\begin{figure}[tb]
\centering
\includegraphics[width= #1]{#2.pdf}
\caption{#3}
\label{f:#4}
\end{figure}}

\newcommand{\todayd}{\the\year/\the\month/\the\day}

\newcommand{\bib}{\bibitem}
\newcommand{\appnum}[1]{\renewcommand{\theequation}{#1.\arabic{equation} }
\setcounter{equation}{0}}

\newcommand{\lb}{\label}
\newcommand{\nt}{\nonumber}
\renewcommand{\fref}[1]{Fig.~\ref{f:#1}}

\newcommand{\bel}{\begin{easylist}}
\newcommand{\eel}{\end{easylist}}
\renewcommand{\bi}[1]{\begin{itemize} #1 \end{itemize}}
\newcommand{\be}[1]{\begin{enumerate} #1 \end{enumerate}}

\def \({\left(}
\def \){\right)}

\def \[{\left[}
\def \]{\right]}
\newcommand{\abs}[1]{\left|#1\right|}

\newcommand{\sumtwo}[2]%
{\mathop{\sum_{#1}}_{#2}}
\newcommand{\sumthree}[3]%
{\mathop{\mathop{\sum_{#1}}_{#2}}_{#3}}
\newcommand{\sumfour}[4]%
{\mathop{\mathop{\mathop{\sum_{#1}}_{#2}}_{#3}}_{#4}} 
\newcommand{\prodtwo}[2]%
{\mathop{\prod_{#1}}_{#2}}
\newcommand{\mintwo}[2]%
{\mathop{\min_{#1}}_{#2}}
\newcommand{\maxtwo}[2]%
{\mathop{\max_{#1}}_{#2}}
\newcommand{\maxthree}[3]%
{\mathop{\mathop{\max_{#1}}_{#2}}_{#3}}
\newcommand{\limtwo}[2]%
{\mathop{\lim_{#1}}_{#2}}
\newcommand{\suptwo}[2]%
{\mathop{\sup_{#1}}_{#2}}
\newcommand{\supthree}[3]%
{\mathop{\mathop{\sup_{#1}}_{#2}}_{#3}}
\newcommand{\supfour}[4]%
{\mathop{\mathop{\mathop{\sup_{#1}}_{#2}}_{#3}}_{#4}} 
\newcommand{\inftwo}[2]%
{\mathop{\inf_{#1}}_{#2}}
\newcommand{\infthree}[3]%
{\mathop{\mathop{\inf_{#1}}_{#2}}_{#3}}
\newcommand{\inffour}[4]%
{\mathop{\mathop{\mathop{\inf_{#1}}_{#2}}_{#3}}_{#4}} 

\newcommand{\bse}{\boldsymbol{e}}

\newcommand{\bsx}{\boldsymbol{x}}
\newcommand{\bsy}{\boldsymbol{y}}

\newcommand{\bsE}{\boldsymbol{E}}

\newcommand{\bbZ}{\mathbb{Z}}
\newcommand{\ep}{\varepsilon}

\newcommand{\pG}{p^{\rm Gibbs}}
\newcommand{\pGb}{p^{\rm G, bath}}
\newcommand{\tlT}{\tilde{T}}
\newcommand{\tlp}{\tilde{p}}
\newcommand{\Emax}{E^{\rm max}}
\newcommand{\Emin}{E^{\rm min}}

\def\rnum#1{\resizebox{0.5em}{\height}{\expandafter{\romannumeral #1}}}
\def\Rnum#1{\resizebox{0.5em}{\height}{\uppercase\expandafter{\romannumeral #1}}}

\begin{document}

\title{Two constructive proofs on d-majorization and thermo-majorization}


\author{Naoto Shiraishi}


\address{
 Department of Physics, Gakushuin University, 1-5-1 Mejiro, Toshima-ku, Tokyo 171-8588, Japan \\
          }
 \ead{naoto.shiraishi@gakushuin.ac.jp}    
 
\begin{indented}
\item \todayd
\end{indented}

\begin{abstract}

Two constructive proofs on d-majorization and thermo-majorization are provided.
In the first part, we present a diagrammatic proof of the equivalence between d-majorization and the existence of a proper stochastic matrix.
We explicitly construct the desired stochastic matrix by using a graphical argument.
In the second part, we present a constructive proof of the equivalence between the Gibbs-preserving map and thermal operation in classical systems.
We construct the desired thermal operation and a heat bath which emulates any Gibbs-preserving map with an arbitrary accuracy.

\end{abstract}

\maketitle

\section{Introduction}

Majorization is an important mathematical research field where we investigate (im)possibility of conversion of probability vectors by some classes of matrices.
The problem of majorization was raised by Hardy, Littlewood and Polya~\cite{HLPbook}.
They prove that a probability vector $p$ majorizes another probability vector $p'$ denoted by $p' \prec p$, which is well visualized by using the Lorenz curve~\cite{Lor, Gour-review}, if and only if there exists a doubly-stochastic matrix $D$ with $p'=Dp$.
There are several proofs on this equivalence~\cite{Ryf, Bhabook}.

The problem of d-majorization, which is raised in various contexts including mathematical statistics (comparison of statistical experiments)~\cite{Bla51, Bla53, Cam}, networks in market\fn{
The name of  {\it d-majorization} is originated in Ref.~\cite{Vei}, where Veinott considers the case of $q=q'=d$.
However, in Veinott's paper, d-majorization is named for a different notion from the present one, and what we now call d-majorization is called {\it d'-majorization} in Ref.~\cite{Vei}.
}~\cite{Vei}, chemical thermodynamics~\cite{RM, RSS}, and purely mathematical interests~\cite{AU, Joe, MOAbook}, consider transitions or comparisons of pairs of probability vectors $(p,q)$.
A pair $(p,q)$ d-majorizes another pair $(p',q')$ denoted by $(p',q') \prec (p,q)$ if and only if there exists a stochastic matrix $T$ with $p'=Tp$ and $q'=Tq$.
Well-known proofs on this fact employ reduction to the problem of majorization~\cite{RSS} or some non-constructive techniques (relying existence theorems or proving by contradiction)~\cite{Bla51, Bla53, Cre, LS}.
If the energy of states are defined and both $q$ and $q'$ are the Gibbs distribution; $q=q'=\pG$, the d-majorization $(p',q') \prec (p,q)$ is also called as thermo-majorization.
Thermo-majorization and other related problems in majorization have been intensively studied as a possible extension of thermodynamics to small systems~\cite{Jan, HHO, Egl, Abe,Bra13, HO, Bra15, FR, Fai, Sag, Mul, Per}, which now becomes a major research field in quantum information theory known as quantum thermodynamics.
Combination of information theory and (d-)majorization bears rich and profound results from entanglement distillation by LOCC~\cite{Nie, Dua}, resource theory of coherence~\cite{DBG, WY} to quantum majorization~\cite{Ren, Bus, BG, Gou}.

In this paper, we provide two constructive proofs on d-majorization and thermo-majorization, both of which have been proven or almost proven in non-constructive ways.
In the first part (Sec.2), we present a diagrammatic construction of the stochastic matrix $T$ for d-majorization.
The conventional proof resorts taking the continuum limit and reducing the problem to majorization~\cite{RSS}.
However, in this approach how to construct the desired stochastic matrix is not shown explicitly, in particular in case of probability vectors with irrational numbers.
In contrast, our approach is irrelevant to the irrationality of probability vectors.
The proposed proof is highly intuitive and the nature of d-majorization becomes clear through this proof.

In the second part (Sec.3), we consider the relation between the Gibbs-preserving map and thermal operation in classical systems.
The Gibbs preserving map (GPM) is a stochastic map which keeps the Gibbs distribution invariant.
The thermal operation (TO) is given by contraction of an energy-conserving map of a composite system of the principal system and a bath system which is initially set as the Gibbs distribution.
These two classes of maps are strongly related.
Any TO is also a GPM, while there exists a GPM which cannot be emulated by TO in the quantum regime~\cite{FOR}.
On the other hand, the GPM and TO are claimed to be equivalent in the classical regime~\cite{HO}.
In the cerebrated paper of Horodecki and Oppenheim~\cite{HO}, a roadmap to emulate GPM by TO is presented.
However, an explicit construction is not shown in this paper (some plausible necessary conditions for a bath system are just listed).
In addition, since their main concern appears to be on the properties after taking the thermodynamic limit of the bath, the relation between GPM and TO with a large but finite size heat bath is not discussed in detail.
In fact, as discussed in Appendix. A, some of the requirement is not fulfilled in a finite size heat bath, and thus the refinement of arguments is inevitable to reach its asymptotic analysis.
In this paper, we clarify the relation between GPM and TO with a large but finite size heat bath.
We first briefly show that there exists a GPM which cannot be emulated by TO exactly, while TO can emulate a restricted class of GPM exactly.
We then demonstrate that any GPM can be emulated by TO with an arbitrary accuracy by explicitly construct the desired bath system and energy-conserving map.
We bound the amount of error (accuracy) rigorously, which tends to zero in the thermodynamic limit.
This evaluation yields the minimum speed of convergence of TO to the given GPM.

\bigskip

Before going to our main results, we here clarify some definitions of words.
Throughout this paper, we consider a linear stochastic map given by a stochastic matrix $T$.
A matrix $T$ is a {\it stochastic matrix} if all matrix elements are nonnegative ($T_{ij}\geq 0$ for any $i,j$) and it satisfies the normalization condition $\sum_i T_{ij}=1$ for any $j$.
A matrix $D$ is a {\it doubly stochastic matrix} if $D$ is a stochastic matrix and $\sum_j D_{ij}=1$ is also satisfied for any $i$.
In this paper, we sometimes say just {\it a map} to refer to the linear stochastic map, and we identify the map itself and the corresponding stochastic matrix.

\section{Proof of the connection between d-majorization and the existence of transition matrix}

\subsection{Lorenz curve and d-majorization}

We first introduce the Lorenz curve.
Consider a pair of probability distributions $(p,q)$ on a system with $n$ states.
For a given pair $(p,q)$, we introduce a permutation $\pi$ on $\{ 1,2,\ldots , n\}$ such that
\eq{
\frac{p_{\pi(1)}}{q_{\pi(1)}}\geq \frac{p_{\pi(2)}}{q_{\pi(2)}}\geq \frac{p_{\pi(3)}}{q_{\pi(3)}}\geq \cdots \geq \frac{p_{\pi(n)}}{q_{\pi(n)}}.
}
We define $p^*$ and $q^*$ by re-ordering the probability distribution $p$ and $q$ as $p^*_i=p_{\pi(i)}$ and $q^*_i=q_{\pi(i)}$.
We introduce the {\it Lorenz curve} of $(p,q)$, which connects point $(0,0)$, $(q^*_1, p^*_1)$, $(q^*_1+q^*_2, p^*_1+p^*_2)$, $( \sum_{i=1}^3q^*_i, \sum_{i=1}^3 p^*_i)$, ... ,$(\sum_{i=1}^{n-1} q^*_i, \sum_{i=1}^{n-1} p^*_i)$, $(\sum_{i=1}^n q^*_i, \sum_{i=1}^n p^*_i)=(1,1)$.
The Lorenz curve of $(p,q)$ is also expressed as the graph of $(x,y)$ given by $y=f(x)$ with
\balign{
f(x)=\max_{{0\leq c_i\leq 1, \ x=\sum_i c_i q_i}} \sum_i c_i p_i . \lb{L-curve-another}
}
If the Lorenz curve of $(p,q)$ lies above that of $(p',q')$ (as \fref{d-maj-setup}), we say that {\it $(p,q)$ d-majorizes $(p', q')$} and write $(p,q) \succ (p',q')$.

It is well known that $(p,q) \succ (p',q')$ holds if and only if there exists a stochastic matrix $T$ such that $p'=Tp$ and $q'=Tq$.
The ``if" part is not difficult.
Let $y=f'(x)$ be the Lorenz curve of $(p',q')$, and $c^*_i$ be the optimum value of $c_i$ in \eref{L-curve-another} at a given $x$ (i.e., $f'(x)=\sum_i c^*_i p'_i$ and $x=\sum_i c^*_i q'_i$ are satisfied).
We then have
\balign{
x&=\sum_i c_i^* q'_i=\sum_{i,j}c_i^* T_{ij}q_j =\sum_j d'_j q_j \\
f'(x)&=\sum_i c_i^* p'_i=\sum_{i,j}c_i^* T_{ij}p_j =\sum_j d'_j p_j 
}
where we defined $d'_j:=\sum_i c_i^*T_{ij}$, which satisfies $0\leq d'_j\leq 1$ owing to the normalization condition $\sum_i T_{ij}=1$.
Hence, we conclude
\eqa{
f'(x)=\sum_j d'_j p_j \leq \max_{{0\leq d_i\leq 1, \ x=\sum_i d_i q_i}} \sum_i d_i p_i =f(x),
}{if-part}
for any $x$, where we defined $f(x)$ as expressing the Lorenz curve of $(p,q)$ by $y=f(x)$.
The relation \eref{if-part} directly implies $(p,q) \succ (p',q')$.

The more difficult part is the ``only if" part:
We should show the existence of a stochastic matrix $T$ such that $p'=Tp$ and $q'=Tq$ under the assumption $(p,q) \succ (p',q')$.
The conventional proof of this equivalence takes the continuum limit of $y$ and reduces the problem to conventional majorization~\cite{RSS}.
Since this proof employs continuum limit, how to construct the desired stochastic matrix $T$ efficiently is still not clear in this proof, in particular for the case that $q$ or $q'$ contain irrational numbers.

\figin{8cm}{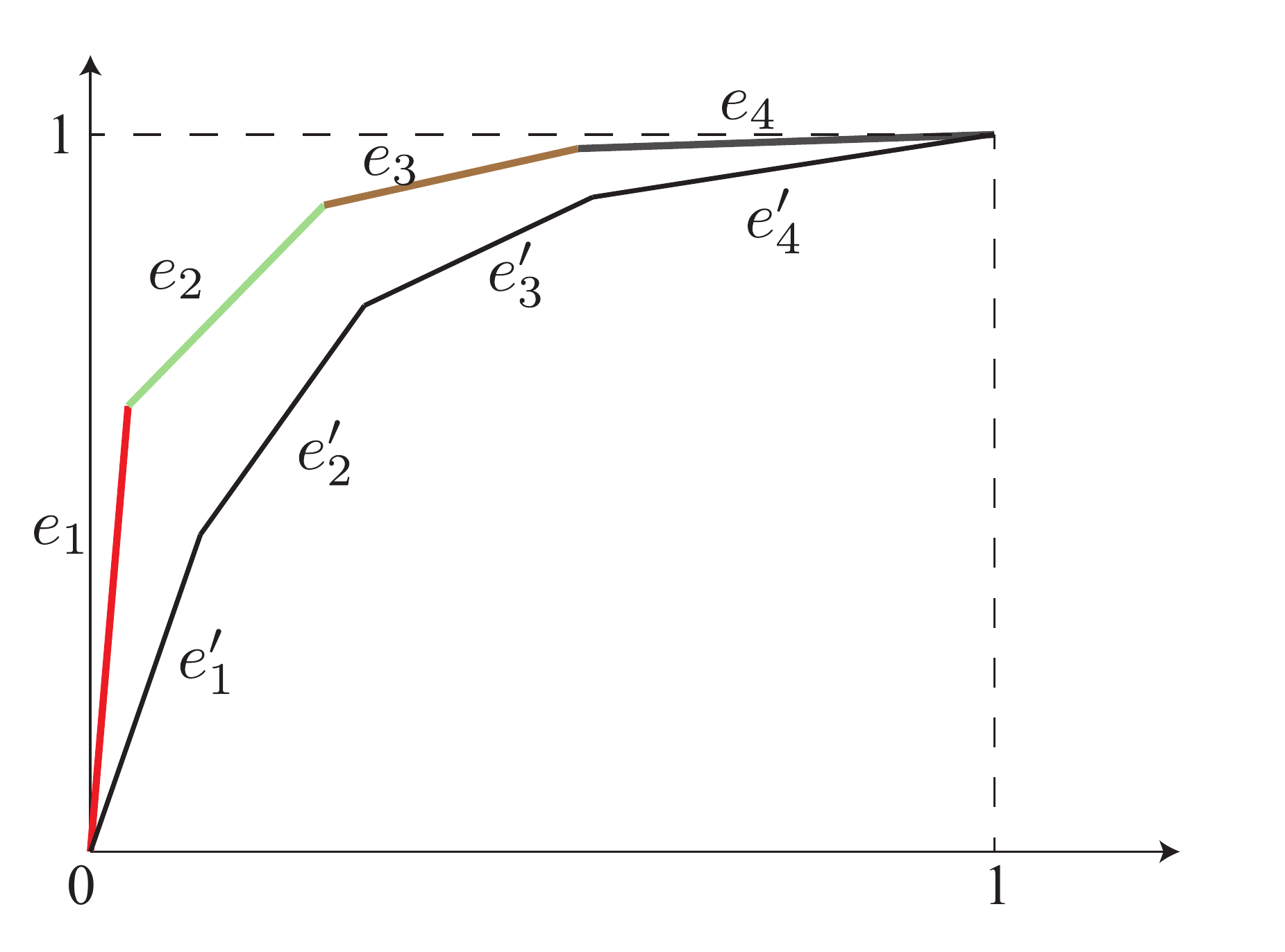}{
An example of the Lorenz curve for the case of $n=4$.
We draw the four edges $e_1$, $e_2$, $e_3$, and $e_4$ as red, light green, brown, and dark gray, respectively.
}{d-maj-setup}

In this section, we demonstrate a very simple and diagrammatic proof of how to construct $T$ from the Lorenz curve explicitly.
Our construction is not a reduction to majorization (we do not use the Birkhoff's theorem), and is irrelevant to the irrationality in the probability distribution.
This proof confirms the existence of the stochastic matrix $T$ graphically.
We first present a diagrammatic proof, and then provide an analytic proof with mathematical induction, which is an analytical expression of the diagrammatic proof.

\subsection{Diagrammatic proof}

By multiplying a proper permutation matrix if needed, without loss of generality we assume
\eq{
\frac{p_1}{q_1}\geq \frac{p_2}{q_2}\geq \cdots \geq \frac{p_n}{q_n}, \hspace{15pt} 
\frac{p'_1}{q'_1}\geq \frac{p'_2}{q'_2}\geq \cdots \geq \frac{p'_n}{q'_n}.
}
To construct the stochastic matrix $T$, we perform the following procedure iteratively.
We denote the Lorenz curves of $(p,q)$ and $(p',q')$ by $C$ and $C'$, respectively.
The edges of $C$ and $C'$ from $(0,0)$ to $(1,1)$ are labeled as $e_1$, $e_2$, ... , $e_n$ and $e'_1$, $e'_2$, ... , $e'_n$.

\figin{16cm}{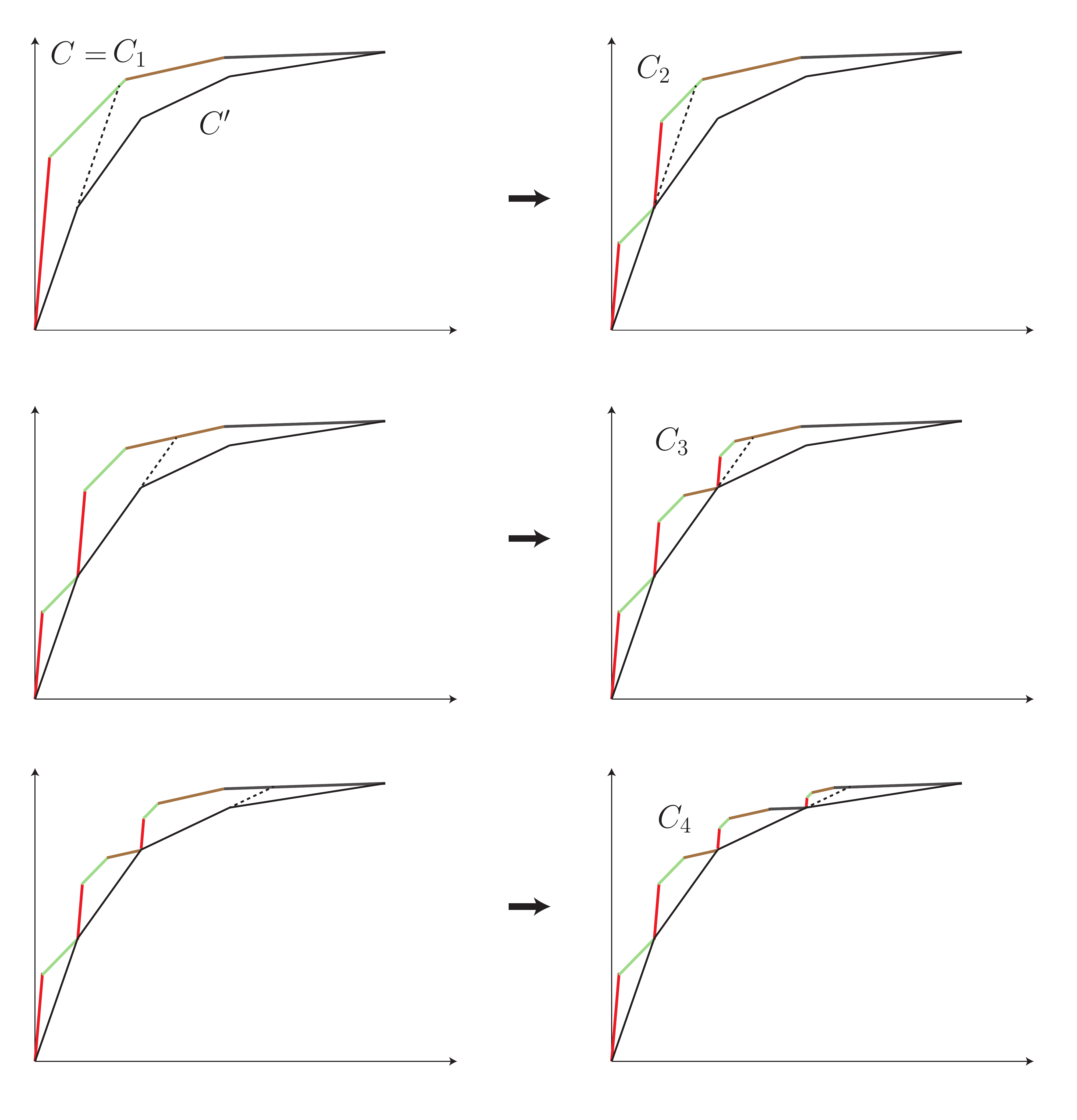}{
The procedure how to construct, $C_1$, $C_2$, $\ldots , C_n$.
(These schematics treat the case of $n=4$ as introduced in \fref{d-maj-setup}.
}{d-maj-step}

We construct $C_1$, $C_2$, ... , $C_n$ recursively.
At the beginning, $C_1$ is set as $C$.
We first construct $C_2$ from $C_1$ and $C'$.
We elongate $e'_1$ to the upper right until it crosses $C_1$ (\fref{d-maj-step}: top-left).
We now modify the curve $C_1$ above the elongated line:
We apply similarity reduction of the curve of $C_1$ above this elongated line to match $e'_1$ and the remaining elongated line, as the first row of \fref{d-maj-step}.
The obtained curve through this procedure is defined as $C_2$.
We note that this modification keeps the total length of red lines and green lines in \fref{d-maj-step} originated from $e_1$ and $e_2$, respectively.

We next construct $C_3$ from $C_2$.
We elongate $e'_2$ to the upper right until it crosses $C_2$.
We then apply similarity reduction of the curve of $C_2$ above this elongated line to match $e'_2$ and the remaining elongated line, as the second row of \fref{d-maj-step}.
The obtained curve is defined as $C_3$.
We apply this procedure $n-1$ times and obtain $C_n$.
Note that the curve $C_n$ touches all corners on $C'$.

\figin{12cm}{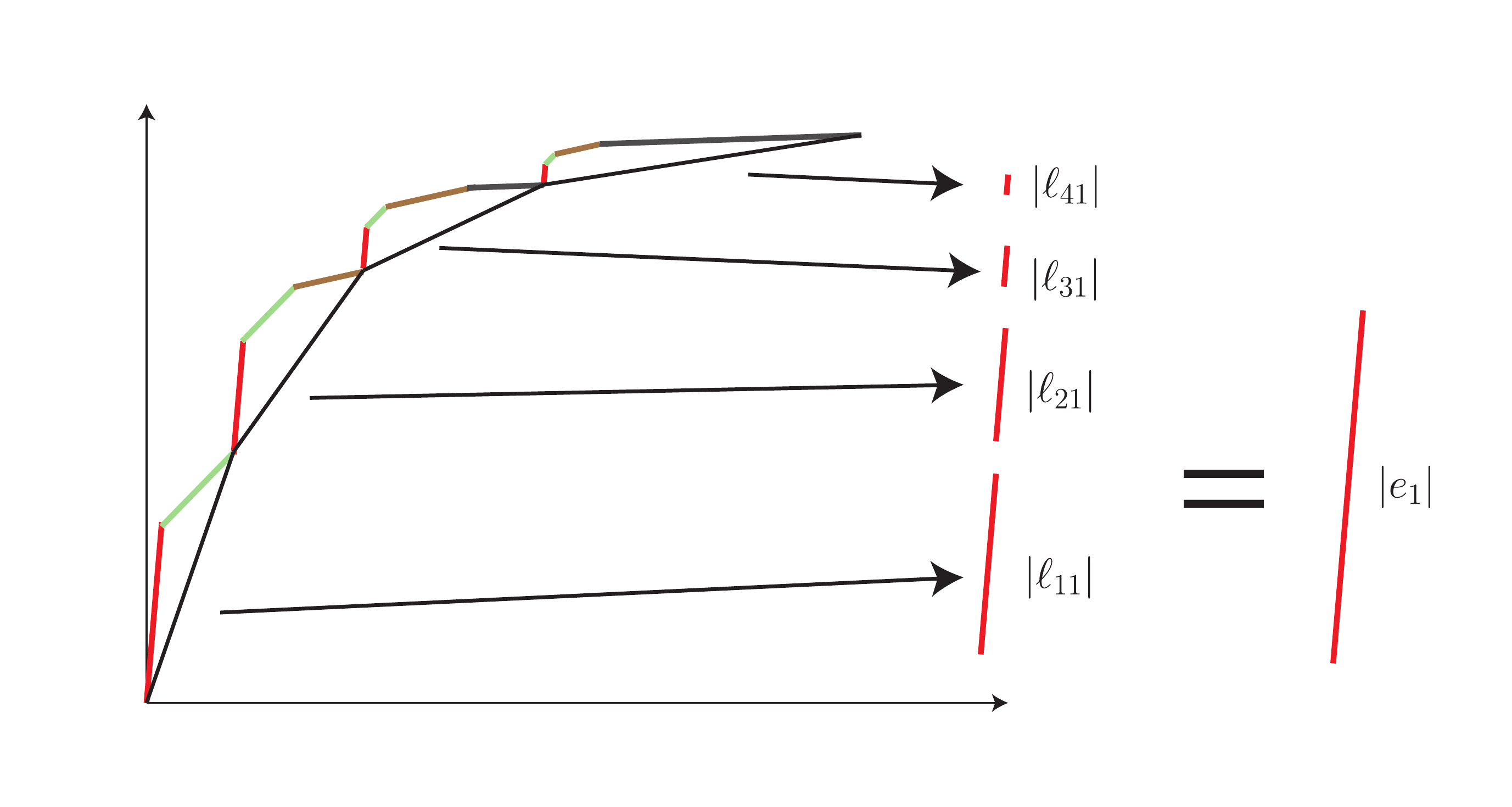}{
Schematic of how to define $T_{j1}=\frac{\abs{\ell_{j1}}}{\sum_{k=1}^n \abs{\ell _{k1}}}=\frac{\abs{\ell_{j1}}}{\abs{e_1}}$ from $C_n$ and $C'$.
}{d-maj-T}

We now construct the stochastic matrix $T$ by using the curves $C_n$ and $C'$.
We demonstrate the construction of $T_{11}, \ldots , T_{n1}$ as an example.
The first column of $T$ corresponds to red lines originated from $e_1$.
We call the red line above $e'_i$ as $\ell_{i1}$.
We then define $T_{i1}$ as
\eq{
T_{i1}:=\frac{\abs{\ell_{i1}}}{\sum_{j=1}^n \abs{\ell _{j1}}}=\frac{\abs{\ell_{i1}}}{\abs{e_1}},
}
where $\abs{\ell}$ represents the length of the line $\ell$.
The denominator $\sum_{j=1}^n \abs{\ell _{j1}}$ is equal to $\abs{e_1}$ because the total length of red lines is invariant.
This definition satisfies the normalization condition $\sum_j T_{j1}=1$.
In a similar manner, $T_{ij}$ is defined as 
\eq{
T_{ij}:=\frac{\abs{\ell_{ij}}}{\sum_{k=1}^n \abs{\ell _{kj}}}=\frac{\abs{\ell_{ij}}}{\abs{e_j}}.
}

\figin{4cm}{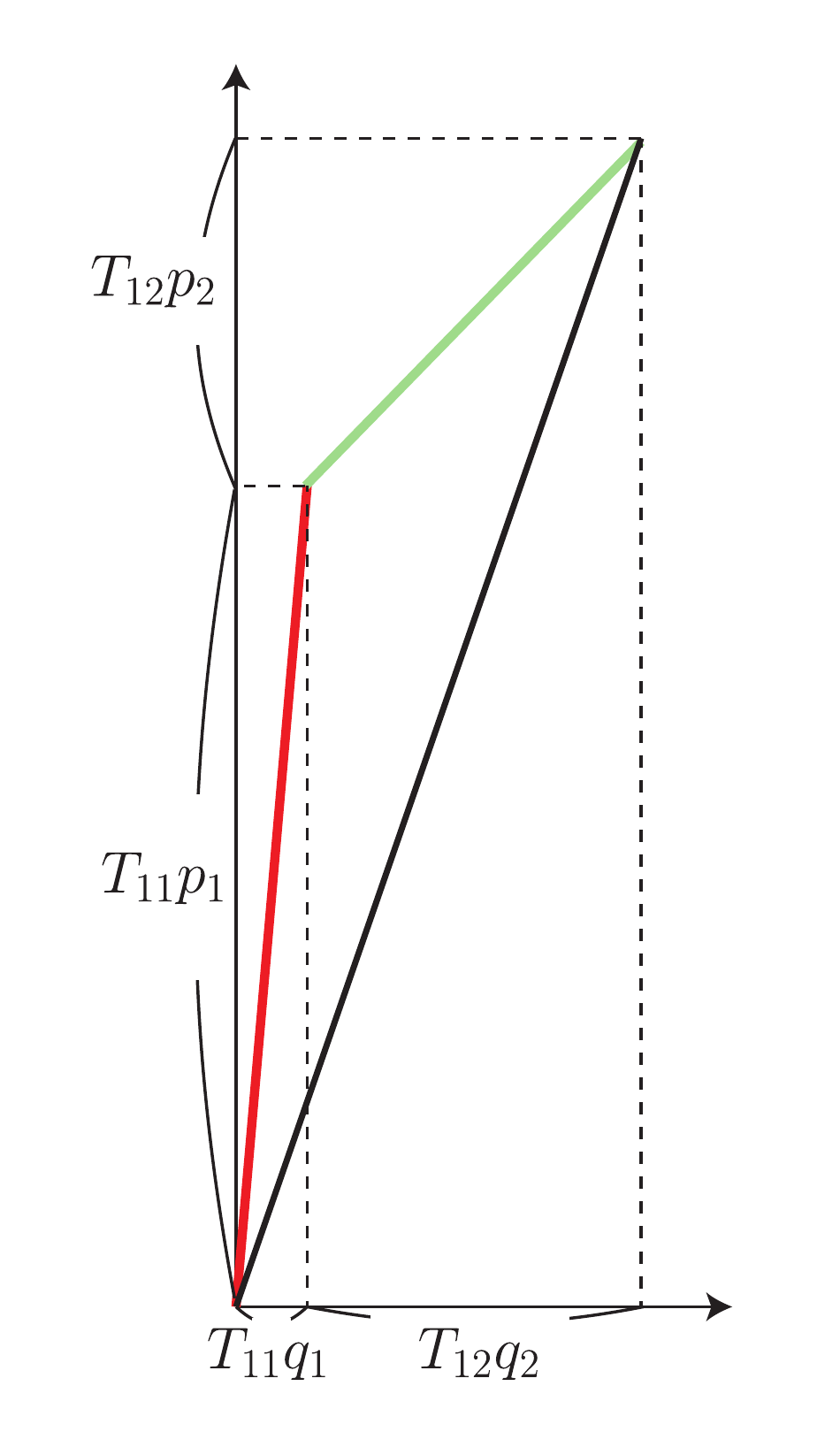}{
We show how the stochastic map $T$ converts $p_i$ and $q_i$ to $p'_1$ and $q'_1$.
}{d-maj-Tp}

We finally demonstrate that this transition matrix $T$ satisfies the desired relations:
\eqa{
p'=Tp, \hspace{10pt} q'=Tq.
}{d-maj-result}
The essence of this fact is shown in \fref{d-maj-Tp}.
The edge $e_1$ can be understood as a vector $(p_1,q_1)$, where we set the direction of vectors as upper right.
Since $\ell_{11}$ is $T_{11}$ times of $e_1$, the vector representation of $\ell_{11}$ is $(T_{11}p_1, T_{11}q_1)$.
In a similar manner, the vector representation of $\ell_{12}$ is $(T_{12}p_2, T_{12}q_2)$.
On the other hand, the vector representation of $e'_1$ (the black line in \fref{d-maj-Tp}) is $(p'_1,q'_1)$.
Comparing these two, we arrive at
\balign{
p'_1&=T_{11}p_1+T_{12}p_2, \\
q'_1&=T_{11}q_1+T_{12}q_2.
}
The case of all other edges $e'_2,\ldots , e'_n$ can be treated in a similar manner, which confirms the desired relations \eref{d-maj-result}.

\subsection{Analytical proof with mathematical induction}

We here provide an analytical proof of the equivalence between d-majorization and the existence of the stochastic matrix $T$.
This proof is essentially the same as the previous diagrammatic proof.

Let $p$ and $q$ be $n$-dimensional vectors, and $p'$ and $q'$ be $m$-dimensional vectors satisfying $\sum_{i=1}^n p_i=\sum_{i=1}^m p'_i\leq 1$ and $\sum_{i=1}^n q_i=\sum_{i=1}^m q'_i\leq 1$.
We suppose that the subscript has already been rearranged in the descending order: 
\eq{
\frac{p_1}{q_1}\geq \frac{p_2}{q_2}\geq \cdots \geq \frac{p_n}{q_n}, \hspace{15pt}
\frac{p'_1}{q'_1}\geq \frac{p'_2}{q'_2}\geq \cdots \geq \frac{p'_m}{q'_m}.
} 
We introduce a {\it generalized Lorenz curve (g-Lorenz curve)} which connects point $(0,0)$, $(q_1, p_1)$, $(q_1+q_2, p_1+p_2)$, $( \sum_{i=1}^3q_i, \sum_{i=1}^3 p_i)$, ... , $( \sum_{i=1}^n q_i, \sum_{i=1}^n p_i)$.
This is a generalization of the conventional Lorenz curve to the case of $\sum_{i=1}^n p_i\neq 1$ and $\sum_{i=1}^n q_i\neq 1$.
If the g-Lorenz curve of $(p,q)$ lies above that of $(p',q')$, we also say that $(p,q)$ d-majorizes $(p', q')$ and write $(p,q) \succ (p',q')$.

We now construct the desired stochastic matrix $T$ and $4m+2$ vectors, $p^1, p^2, \ldots , p^{m+1}, q^1,\ldots , q^{m+1}$ and $p'^1, \ldots p'^m, q'^1, \ldots , q'^m$.
We first set $p^1=p$, $q^1=q$, $p'^1=p'$, and $q'^1=q'$.
We shall construct $p^{k+1},q^{k+1}, p'^{k+1},q'^{k+1}$ from $p^k, q^k, p'^k, q'^k$ recursively.
As shown below, through this construction $(p^k,q^k) \succ (p'^k,q'^k)$ is satisfied for any $k$.

\figin{14cm}{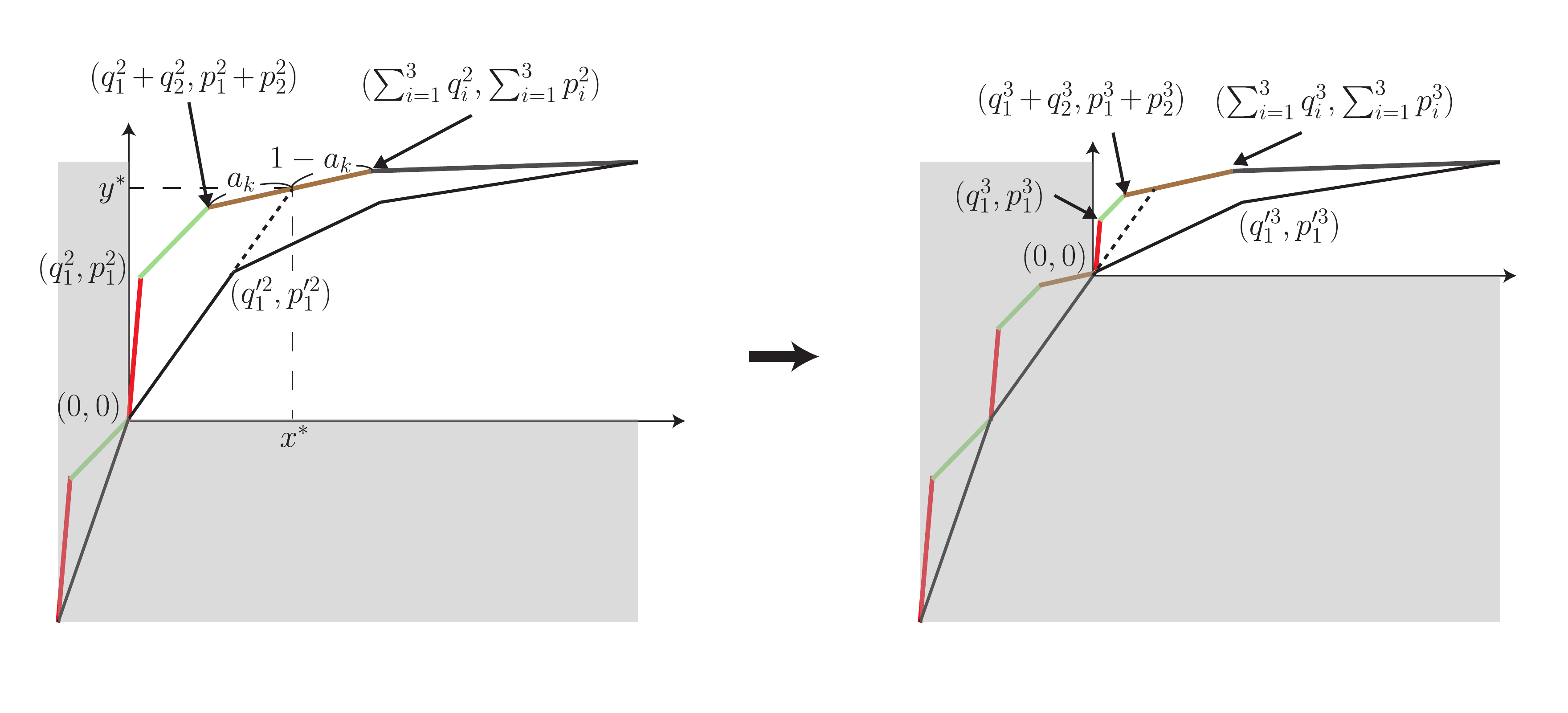}{
An example of the correspondence between analytic expression and diagrammatic expression.
}{d-maj-analytic}

Due to the condition $(p^k,q^k) \succ (p'^k,q'^k)$, a line $y=\frac{p'^k_1}{q'^k_1}x$ crosses the g-Lorenz curve of $(p^k,q^k)$ other than the origin $(0,0)$.
We denote this point by $(x^*, y^*)$ and let $c_k$ be an integer that the point $(x^*, y^*)$ settles between $( \sum_{i=1}^{c_k} q^k_i, \sum_{i=1}^{c_k} p^k_i)$ and $( \sum_{i=1}^{c_k+1} q^k_i, \sum_{i=1}^{c_k+1} p^k_i)$.
In other words, there exists $0\leq a_k \leq 1$ such that
\balign{
x^*&=\sum_{i=1}^{c_k} q^k_i+a_k q^k_{c_k+1} \\
y^*&=\sum_{i=1}^{c_k} p^k_i+a_k p^k_{c_k+1}
}
and $y^*=\frac{p'^k_1}{q'^k_1}x^*$ (see \fref{d-maj-analytic}).
In case of $p^k_1/q^k_1=p'^k_1/q'^k_1$ (i.e., the line $y=\frac{p'^k_1}{q'^k_1}x$ touches the first edge of the g-Lorenz curve of $(p^k,q^k)$), we set $(x^*,y^*)=(p'^k_1, q'^k_1)$.
We note that $(p^k,q^k) \succ (p'^k,q'^k)$ implies $p'^k_1\leq x^*$.
We then construct $n$-dimensional vectors $p^{k+1},q^{k+1}$ and $m-k$-dimensional vectors $p'^{k+1},q'^{k+1}$ as
\balign{
p^{k+1}_i&=\cases{
\( 1-\frac{p'^k_1}{y^*}\) p^k_i & $i\leq c_k$ \\
\( 1-\frac{p'^k_1a_k}{y^*}\)  p^k_{c_k+1} & $i=c_k+1$ \\
p^k_i &$i\geq c_k+2$
} \\
q^{k+1}_i&=\cases{
\( 1-\frac{p'^k_1}{y^*}\) q^k_i & $i\leq c_k$ \\
\( 1-\frac{p'^k_1a_k}{y^*}\)  q^k_{c_k+1}  & $i=c_k+1$ \\
q^k_i &$i\geq c_k+2$
}
}
and
\balign{
p'^{k+1}_i&=p'^k_{i+1} \\
q'^{k+1}_i&=q'^k_{i+1}.
}
By construction, 
\eq{
\frac{p^{k+1}_i}{q^{k+1}_i}=\frac{p^k_i}{q^k_i}
}
and
\balign{
\sum_{i=1}^s p^{k+1}_i &=\( \sum_{i=1}^s p^k_i\) -p'^k_1 \\
\sum_{i=1}^s q^{k+1}_i &=\( \sum_{i=1}^s q^k_i\) -q'^k_1
}
with any $s\geq c_k+1$ are satisfied, which indicate $(p^{k+1}, q^{k+1})\succ (p'^{k+1}, q'^{k+1})$ for the following reason.
Let $y=f_{k+1}(x)$ and $y=f'_{k+1}(x)$ be the Lorenz curves of $(p^{k+1}, q^{k+1})$ and $(p'^{k+1}, q'^{k+1})$.
Then, for $x\leq x^*-p'^k_1$, we have $f'_{k+1}(x)\leq (p'^k_1/q'^k_1)x\leq f_{k+1}(x)$.
For $x>x^*-p'^k_1$, both $f_{k+1}(x)=f_k(x+p'^k_1)$ and $f'_{k+1}(x)=f'_k(x+p'^k_1)$ are satisfied, and hence $(p^k,q^k)\succ (p'^k,q'^k)$ (i.e., $f'_k(x)\leq f_k(x)$) implies $f'_{k+1}(x)\leq f_{k+1}(x)$.
In conclusion, $f'_{k+1}(x)\leq f_{k+1}(x)$ holds for any $x$, which is equivalent to $(p^{k+1}, q^{k+1})\succ (p'^{k+1}, q'^{k+1})$

We now construct the desired stochastic matrix $T$ as
\balign{
T_{ij}:=\frac{p^i_j-p^{i+1}_j}{p^1_j} =\frac{q^i_j-q^{i+1}_j}{q^1_j},
}
which realizes desired relations:
\balign{
\sum_j T_{ij}p_j=&\sum_j (p^i_j-p^{i+1}_j)=\sum_j p'^i_j-\sum_j p'^{i+1}_j=p'^i_1=p'_i, \\
\sum_j T_{ij}q_j=&\sum_j (q^i_j-q^{i+1}_j)=\sum_j q'^i_j-\sum_j q'^{i+1}_j=q'^i_1=q'_i.
}

\section{Proof of emulation of Gibbs-preserving map by thermal operation}

\subsection{Definitions and claims}

In this section, we consider emulations of a map $A$ by some classes of maps $S$.
We here clarify some definitions which we shall use in this paper:

\

\bi{
\item A class of maps $S$ emulates a map $A$ {\it exactly} if there exists a map $B\in S$ such that $A_{ij}=B_{ij}$ for any $i,j$.
\item A class of maps $S$ emulates a map $A$ {\it with an arbitrary accuracy} if for any $\ep>0$ there exists a map $B\in S$ such that $\abs{A_{ij}-B_{ij}}<\ep$ for any $i,j$.
\item A class of maps $S$ realizes a state conversion from $p$ to $p'$ {\it exactly} if there exists a map $B\in S$ such that $\sum_jB_{ij}p_j=p'_i$ for any $i$.
}

\

We introduce two important classes of maps, the Gibbs-preserving map (GPM) and thermal operation (TO).
Through defining these maps, we implicitly fix an inverse temperature $\beta$.
Let $\pG_i:=e^{-\beta E_i}/Z$ be a Gibbs distribution, where $E_i$ is the energy of the state $i$ and $Z$ is a constant for normalization.
A map $T$ is a {\it Gibbs-preserving map} if the Gibbs distribution $\pG$ is invariant under this map: $\pG=T\pG$.

To define a thermal operation, we introduce a bath system, and consider a composite system of the (principal) system and the bath.
Throughout this paper, we consider a large but finite size heat bath, not an infinite size heat bath.
We denote by $(i,a)$ the state of the composite system where the state of the system is $i$ and that of the bath is $a$.
A map on the composite system $D$ is {\it energy conserving} if $D_{(i,a),(j,b)}$ takes a nonzero value only when $E_i+E_a=E_j+E_b$ is satisfied.
A map on the system $T$ is a {\it thermal operation} if there exists a proper bath system and an energy-conserving doubly-stochastic map on the composite system $D$ such that $T_{ij}$ is given by
\eq{
T_{ij}=\sum_{a,b}D_{(i,a),(j,b)}\pGb_b,
}
where $\pGb$ is the Gibbs distribution of the bath.
Note that in quantum systems thermal operation is defined as the classical mixture of $\Lambda(\rho)=\Tr_{\rm B}[U_{\rm SB}(\rho\otimes \tau)U_{\rm SB}^\dagger]$, where $U_{\rm SB}$ is an energy-conserving unitary operator and $\tau$ is the Gibbs state of the bath.
The classical counterpart of unitary operators is the permutation matrix, and Birkhoff's theorem suggests that the mixture of permutation matrices is equivalent to the doubly-stochastic matrix~\cite{Bhabook}.
Thus, the above definition is a classical counterpart of the conventional definition of thermal operation in quantum thermodynamics.

\bigskip

In this section, we clarify the relation between the Gibbs-preserving map (GPM) and thermal operation (TO) in classical systems.
We claim that
\be{
\item There exists a pair of states $p,p'$ such that GPM realizes the state conversion from $p$ to $p'$ exactly while TO cannot.
\item TO emulates any GPM with small off-diagonal elements exactly.
\item TO emulates any GPM with an arbitrary accuracy.
}
Our main achievement in this section is the claim 3, emulation of GPM by TO with an arbitrary accuracy.
We explicitly construct the desired TO for any given GPM.

The celebrated paper by Horodecki and Oppenheim~\cite{HO} presents the basic idea how to construct TO emulating a given (semi-)classical GPM.
However, their description is not constructive, and some of their requirement is not fulfilled in a large but finite size bath.
Thus, what bath indeed works and how to construct a correct heat bath are not fully clarified.
This point is discussed in detail in Appendix. A.
Related to this point, since the main interest of Horodecki and Oppenheim seems to be on infinite size heat baths after taking the thermodynamic limit, the question how large the heat bath should be to achieve the given accuracy of approximation has not yet been addressed.
To answer these questions, we explicitly construct TO with a large but finite size bath emulating a given GPM.
Our construction accompanies the evaluation of the speed of convergence, and clarifies how large the heat bath should be for achieving a given accuracy of approximation.

\subsection{State conversion realized by GPM but not by TO}

Consider a two-level system with states up and down with energy difference $E$.
We claim that the map from $p$ to $p'$ with
\eq{
p=\( \begin{array}{c}1\\0 \end{array}\) , \hspace{10pt} 
p'=\( \begin{array}{c}0\\1 \end{array}\) 
}
is realized by GPM but not by TO with finite temperature exactly.

This map is realized by the following GPM:
\eq{
T=\( \begin{array}{cc}
0&e^{-\beta E} \\
1& 1-e^{-\beta E}
\end{array}
\) .
}
On the other hand, no TO realizes this map exactly.
The reason is as follows.
This map requires the transition from up to down with probability 1.
However, due to the energy conservation law, if the state in the bath has the highest energy, the transition from up to down in the system never occurs.
The bath takes the state with the highest energy with finite probability, which prevents exact realization of this map by TO.

\subsection{TO exactly emulates GPM with small off-diagonal elements}

In the previous subsection, we construct a counterexample to exact emulation of GPM by TO by using large off-diagonal elements.
On the other hand, if all the off-diagonal elements of GPM is sufficiently small, there exists TO which emulates this GPM exactly.

Consider a $d$-state system with states $\{ 0,1,\ldots , d-1\}$ and their energies $E_0=0\leq E_1\leq E_2 \leq \cdots \leq E_{d-1}$.
Given a GPM $T$ with small off-diagonal elements.
The detailed condition for smallness of off-diagonal elements is provided later.

We now construct a bath and a doubly-stochastic matrix on the composite system.
The bath is also a $d$-state system with states $\{ 0,1,\ldots , d-1\}$ and the energy of the state $i$ is set to $-E_i$.
We denote by $(i,j)$ the state of the composite system where the state of the system is $i$ and that of the bath is $j$.
We introduce a doubly-stochastic matrix on the composite system $D_{(k,l)(i,j)}$, whose matrix elements take nonzero values only if $k=l$ and $i=j$.
Henceforth, we abbreviate $D_{(i,i)(j,j)}$ as $D_{ij}$.

Using $T_{ij}$, we set $D_{ij}$ ($i\neq j$)  as
\eq{
D_{ij}:=T_{ij}\pG_j Z\bar{Z}, 
}
where $\pG_j$ is the Gibbs distribution of the system
\eq{
\pG_j:=\frac{e^{-\beta E_j}}{Z},
}
and $Z$ and $\bar{Z}$ are respectively distribution functions of the system and the bath:
\balign{
Z&:=\sum_i e^{-\beta E_i}, \\
\bar{Z}&:=\sum_i e^{\beta E_i}.
}
By construction, the dynamics of $D$ contracted to the system reproduces the dynamics given by $T$:
\eq{
D_{ij}\frac{e^{\beta E_j}}{\bar{Z}}=T_{ij}.
}

We now clarify the condition of the smallness of the off-diagonal elements of $T$.
We require that $T$ satisfies
\eq{
\sum_{i(\neq j)}T_{ij}\pG_j Z\bar{Z} \leq 1
}
for any $j$.
Under this requirement, the diagonal elements of $D$ given by
\eq{
D_{jj}:=1-\sum_i D_{ij}
}
satisfy the nonnegativity condition $D_{jj}\geq 0$.
In addition, since $T$ is a Gibbs-preserving map, we find that the matrix $D$ is a doubly-stochastic matrix:
\eq{
\sum_j D_{ij}=1+ \sum_{j(\neq i)} (D_{ij}- D_{ji})=1+  Z\bar{Z} \sum_{j(\neq i)}(T_{ij}\pG_j-T_{ji}\pG_i)=1.
}

\subsection{Emulation of general GPM by TO with an arbitrary accuracy}

In this subsection, we construct a TO which emulates any given GPM with an arbitrary accuracy $(1>)\ep>0$.
Consider a $d$-state system with states $\{ 0,1,\ldots , d-1\}$ and their energies $E_0=0\leq E_1\leq E_2 \leq \cdots \leq E_{d-1}$.
Given a GPM $T$, we construct a TO emulating $T$ with an accuracy $\ep$.

We construct the state space of the bath as follows:
The state of the bath is characterized by a pair $(\bsx, m)$ where $\bsx$ is a $d-1$-dimensional lattice point $\bsx=(x_1,x_2,\ldots , x_{d-1})$ with integers $x_i\in \{ 0,1, 2,\ldots, L\}$ and $m$ is a label to distinguish degenerated states $m\in \{ 1,2,\ldots N(\bsx)\}$.
A state with the lattice point $\bsx$ has the energy $-\sum_i E_ix_i$.
The number $N(\bsx)$ represents the degree of degeneracy with $\bsx$.
We call the set of states at a certain lattice point as {\it island}.
A single lattice point corresponds to an island.

For a given accuracy $\ep>0$, we set $L$ as satisfying
\eqa{
\( \frac{L}{L+1}\) ^{d-1}>1-\frac{\ep}{16},
}{cond-L}
which is fulfilled when $L>16(d-1)/\ep$ is satisfied.
We set the number of states at island $N(\bsx)$ in order that in the Gibbs distribution all islands appear with almost the same probability.
Toward this goal, we set $N(\bsx)$ as satisfying
\eqa{
\abs{\frac{N(\bsx)e^{\beta \bsE \cdot \bsx}}{\sum_{\bsx'} N(\bsx')e^{\beta \bsE \cdot \bsx'} }\cdot (L+1)^{d-1}-1}<\frac{\ep}{32},
}{ratio-approx}
where $\bsE\cdot \bsx:=\sum_i E_i x_i$ is the ordinal inner product.
For any $\ep>0$ there exists a set of $N(\bsx)$ satisfying \eref{ratio-approx} because the rational number is dense.

\figin{10cm}{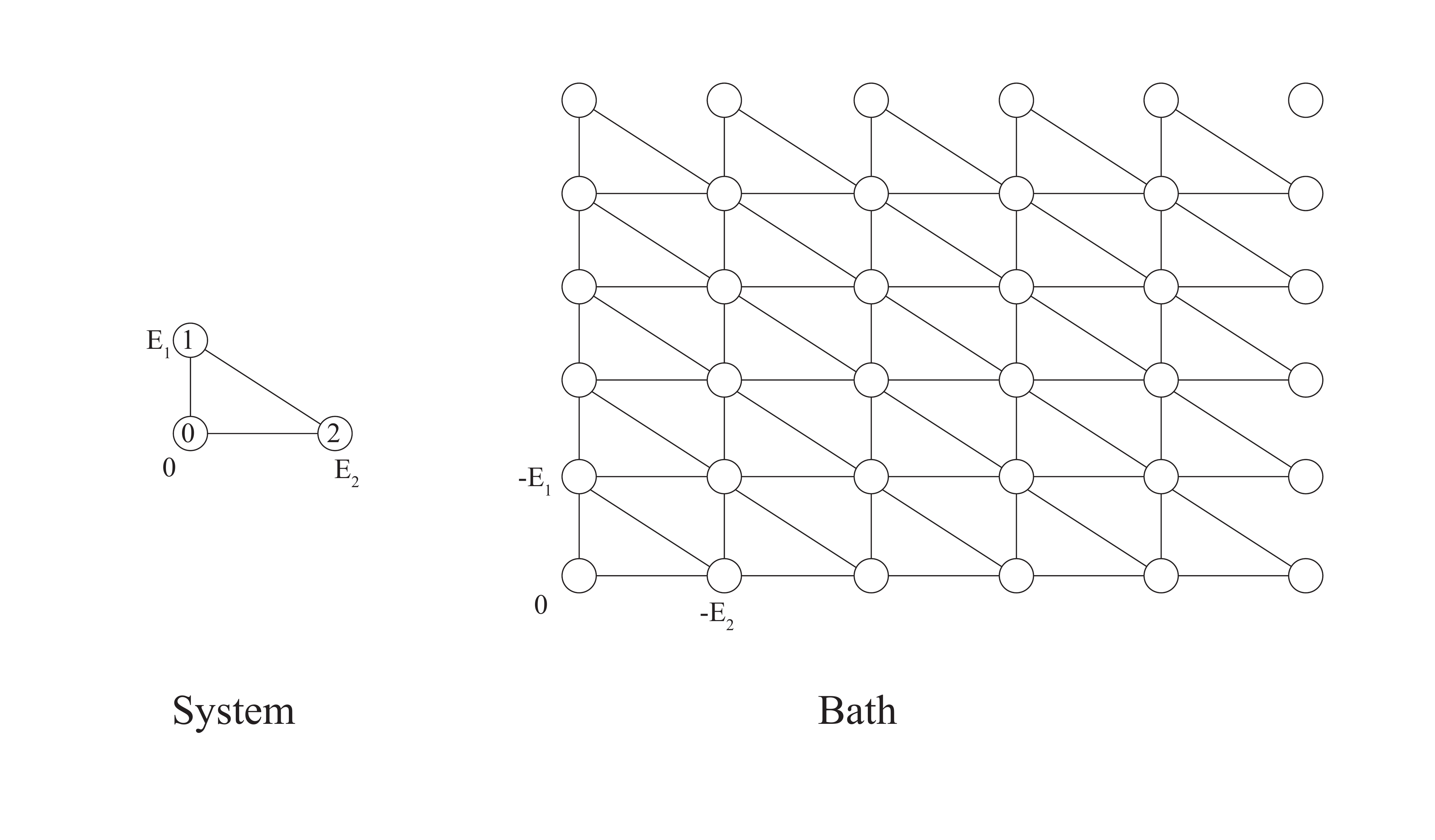}{
An example of the state space of the system (Left: case of $d=3$), and the corresponding state space of the bath (Right).
We remark that a single circle in the state space of the bath (right figure) represents a set of states (island) of the bath with the same energy, not a single state of the bath.
}{setup-TO}

Suppose that the initial state of the system is $i$, and the initial island of the bath is $\bsx=(x_1,x_2,\ldots , x_{d-1})$.
We express the state-island pair with a square bracket as $[i,\bsx]$.
Meanwhile, we forget the micro states in islands and regard an island as a single state of the bath.
Let $\bse_n$ be the $d-1$ dimensional unit vector (The $n$-th element is 1, and others are 0. $\bse_0$ is set to the zero vector), and we define $\bsx^{i, \bsy}:=\bsy+\bse_i$ ($i\in \{ 0,1,\ldots , d-1\}$).
Then, all pairs $[j, \bsx^{j, \bsy}]$ with the same $\bsy$ take the same energy, which can be converted to each other through energy-conserving maps.
An example is drawn in \fref{TO-corresp}.
If the initial state is the pair of the gray state and island, the bath system can take states with solid lines.

\bigskip

\figin{10cm}{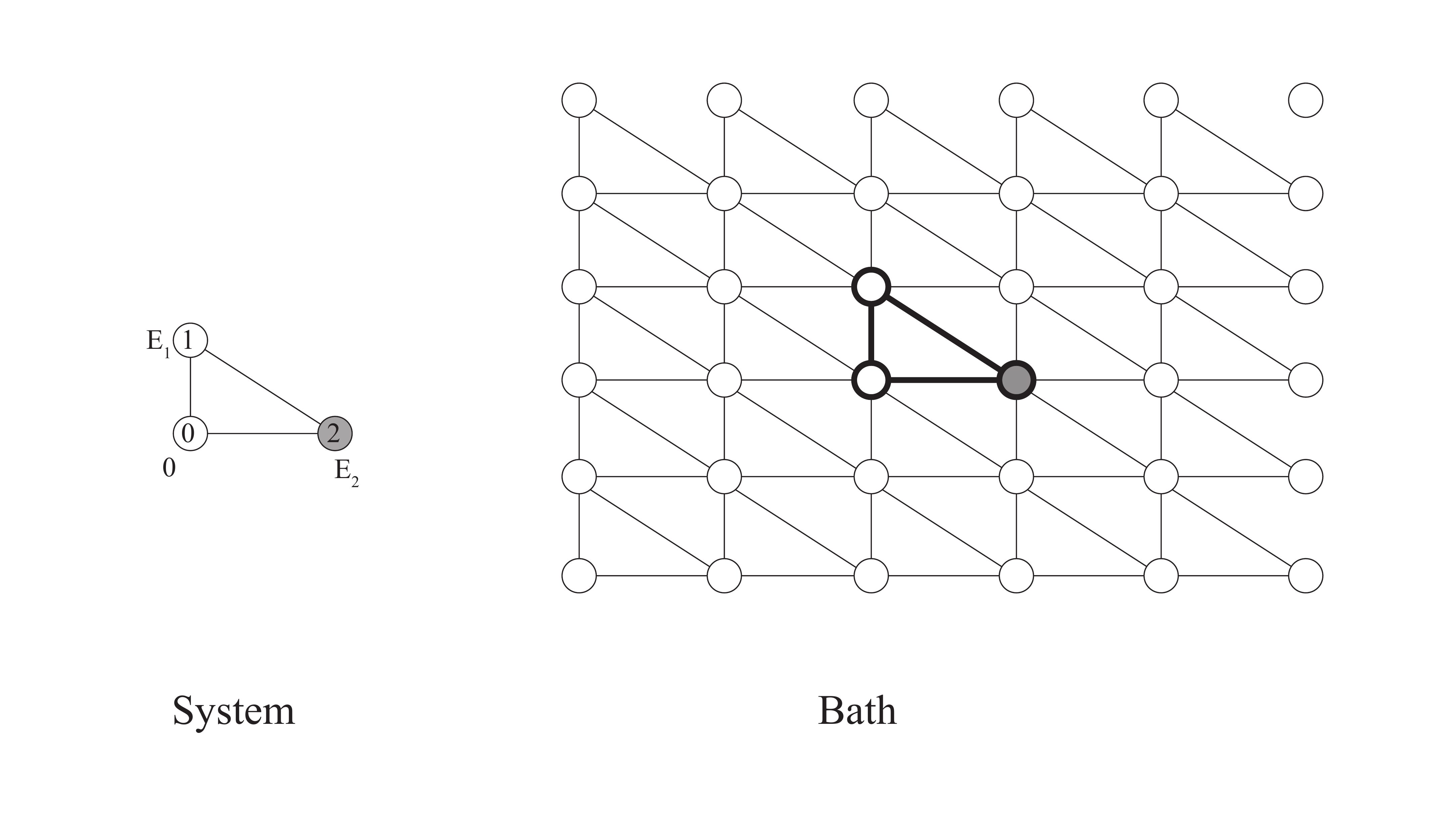}{
If the initial state of the system and the initial island of the bath are the gray circles, the composite system can take only the states and islands on the solid cones.
In addition, the system and the bath take the same vertices on the two cones due to the law of energy conservation.
}{TO-corresp}

In description with $[j, \bsx^{j, \bsy}]$, an energy-conserving map keeps $\bsy$ invariant, and converts only the variable $j$ ($j\in \{ 0,1,\ldots ,d-1\}$)\fn{
Of course, if energies take rational numbers, there exists accidental coincidence that two pairs with different $\bsy$ have the same energy.
However, in this case we set all transition probabilities between pairs with different $\bsy$s as zero.
}.
In other words, the state space is decomposed into subsets of states with the same $\bsy$ (states with the same energy), and state conversions occur only inside the subsets.
On the basis of this observation, we fix $\bsy$ and construct a $\bsy$-dependent transition matrix $\tlT^{\bsy}$ on the system, which is close to the given GPM, $T$, and whose stationary distribution is given by
\eq{
\tlp^{\bsy} _i:=\frac{N(\bsx^{i, \bsy})}{\sum_j N(\bsx^{j, \bsy})}.
}
We fix $\bsy$, and hereafter we drop the superscript $\bsy$ for visibility.
Note that the stationary distribution $\tlp$ is a rational number approximation of the Boltzmann distribution of the transition matrix $T$ defined as $\pG_i:=e^{-\beta E_i}/Z$.
Since the condition \eref{ratio-approx} is equivalent to
\eq{
\( 1-\frac{\ep}{32} \) C <N(\bsx)e^{\beta \bsE \cdot \bsx} < \( 1+\frac{\ep}{32} \) C
}
with $C:= ( \sum_{\bsx'} N(\bsx')e^{\beta \bsE \cdot \bsx'})/(L+1)^{d-1}$, we obtain
\eqa{
\( 1-\frac{\ep}{16}\) \pG_i \leq \frac{1-\frac{\ep}{32}}{1+\frac{\ep}{32}}\pG_i \leq \tlp_i \leq \frac{1+\frac{\ep}{32}}{1-\frac{\ep}{32}} \pG_i \leq \( 1+\frac{\ep}{8}\) \pG_i ,
}{tlp-1}
and in a similar manner to above we obtain
\eqa{
\( 1-\frac{\ep}{16}\) \tlp_i \leq \pG_i \leq \( 1+\frac{\ep}{8}\) \tlp_i .
}{tlp-2}

We now define $\tlT_{ij}$ ($i\neq j$) as
\eqa{
\tlT_{ij}:=\(1-\frac{\ep}{4}\) \( T_{ij}-\frac{(T_{ij}\tlp_j-T_{ji}\tlp_i)- (T_{ij}\pG_j-T_{ji}\pG_i)}{\tlp_i+\tlp_j} \) .
}{tlT-def}
By construction, the stationary probability current from $j$ to $i$ under $\tlT$ is $1-\ep/4$ times of the stationary probability current under $T$: 
\eq{
(\tlT_{ij}\tlp_j-\tlT_{ji}\tlp_i)=\( 1-\frac{\ep}{4}\) (T_{ij}\pG_j-T_{ji}\pG_i).
}
The multiplication of $1-\ep/4$ is necessary in the later evaluation \eref{tlT-diag}.
In order to satisfy the normalization condition, we define the diagonal elements of $\tlT$ as
\eq{
\tlT_{jj}:=1-\sum_{i(\neq j)} \tlT_{ij}.
}

We first check that $\tlT$ is indeed a transition matrix (i.e., $\tlT_{jj}$ is nonnegative).
The sum of the second term of \eref{tlT-def} over $i$ is bounded from above as
\balign{
&-\sum_{i(\neq j)}\frac{(T_{ij}\tlp_j-T_{ji}\tlp_i)- (T_{ij}\pG_j-T_{ji}\pG_i)}{\tlp_i+\tlp_j} \nt \\
=&-\sum_{i(\neq j)}\frac{T_{ij}(\tlp_j-\pG_j)}{\tlp_i+\tlp_j}+\sum_{i(\neq j)}\frac{T_{ji}(\tlp_i-\pG_i)}{\tlp_i+\tlp_j} \nt \\
\leq& \sum_{i(\neq j)}\frac{T_{ij}\tlp_j }{\tlp_i+\tlp_j} \frac{\ep}{8}+\sum_{i(\neq j)}\frac{T_{ji}\pG_i}{\tlp_i+\tlp_j} \frac{\ep}{8} \nt \\
\leq& \sum_{i(\neq j)}T_{ij} \frac{\ep}{8}+\sum_{i(\neq j)}\frac{T_{ji}\pG_i}{\tlp_j} \frac{\ep}{8} \nt \\
=&\sum_{i(\neq j)}T_{ij} \frac{\ep}{8}+\sum_{i(\neq j)}\frac{T_{ij}\pG_j}{\tlp_j} \frac{\ep}{8} \nt \\
\leq& \( 2+\frac{\ep}{8}\) \frac{\ep}{8}\sum_{i(\neq j)}T_{ij} ,
}
where we used $\sum_{i(\neq j)} T_{ij}\pG_j=\sum_{i(\neq j)} T_{ji}\pG_i$,  the stationary condition for $T$.
Hence, by inserting the above inequality to \eref{tlT-def}, the nonnegativity of $T$ ($\sum_{i(\neq j)}T_{ij}\leq 1$) leads to the nonnegativity of $\tlT$:
\balign{
\sum_{i(\neq j)}\tlT_{ij}\leq &\(1-\frac{\ep}{4}\) \[ 1+\frac{\ep}{4}+ \frac{\ep^2}{64}\] \sum_{i(\neq j)}T_{ij} \leq \sum_{i(\neq j)}T_{ij}\leq 1. \lb{tlT-diag}
}

We next demonstrate that $\tlT$ is close to $T$.
In case of $i\neq j$, by using $0\leq T_{ij}, T_{ji}\leq 1$ and Eqs.~\eref{tlp-1} and \eref{tlp-2}, the difference between $\tlT_{ij}$ and $T_{ij}$
\eq{
\tlT_{ij}-T_{ij}=-\frac{\ep}{4}T_{ij}-\(1-\frac{\ep}{4}\) \frac{(T_{ij}\tlp_j-T_{ji}\tlp_i)- (T_{ij}\pG_j-T_{ji}\pG_i)}{\tlp_i+\tlp_j} 
}
is bounded from both above and below as
\balign{
\tlT_{ij}-T_{ij}&\leq \(1-\frac{\ep}{4}\) \[ \frac{T_{ij}\tlp_j }{\tlp_i+\tlp_j} \frac{\ep}{8}+\frac{T_{ji}\tlp_i}{\tlp_i+\tlp_j} \frac{\ep}{16}\] \leq \frac{3\ep}{16}, \\
\tlT_{ij}-T_{ij}&\geq -\frac{\ep}{4}-\(1-\frac{\ep}{4}\) \[ \frac{T_{ij}\tlp_j }{\tlp_i+\tlp_j} \frac{\ep}{16}+\frac{T_{ji}\tlp_i}{\tlp_i+\tlp_j} \frac{\ep}{8}\]  \geq -\frac{7\ep}{16}.
}
In case of $i=j$, we employ the relation $T_{ii}=1-\sum_{j(\neq i)}T_{ji}$ and  $\tlT_{ii}=1-\sum_{j(\neq i)}\tlT_{ji}$, and evaluate $\sum_{j(\neq i)}T_{ji}$ and $\sum_{j(\neq i)}\tlT_{ji}$.
The bound from below has already been obtained in \eref{tlT-diag}: $0\leq \tlT_{jj}-T_{jj}$.
To obtain the bound from above, we follow a similar calculation for $i\neq j$ and arrive at
\balign{
&-\sum_{i(\neq j)}\frac{(T_{ij}\tlp_j-T_{ji}\tlp_i)- (T_{ij}\pG_j-T_{ji}\pG_i)}{\tlp_i+\tlp_j} \nt \\
=&-\sum_{i(\neq j)}\frac{T_{ij}(\tlp_j-\pG_j)}{\tlp_i+\tlp_j}+\sum_{i(\neq j)}\frac{T_{ji}(\tlp_i-\pG_i)}{\tlp_i+\tlp_j} \nt \\
\geq& -\[ \sum_{i(\neq j)}\frac{T_{ij}\tlp_j }{\tlp_i+\tlp_j} \frac{\ep}{16}+\sum_{i(\neq j)}\frac{T_{ji}\pG_i}{\tlp_i+\tlp_j} \frac{\ep}{16}\] \nt \\
\geq& -\( 2+\frac{\ep}{8}\) \frac{\ep}{16}\sum_{i(\neq j)}T_{ij},
}
which implies 
\balign{
&\sum_{i(\neq j)}\tlT_{ij}\geq \( 1-\frac{\ep}{4}\) \sum_{i(\neq j)}T_{ij}-\(1-\frac{\ep}{4}\) \( 2+\frac{\ep}{8}\) \frac{\ep}{16}\sum_{i(\neq j)}T_{ij} \nt \\
 \geq& \( 1 -\frac{3\ep}{8}\) \sum_{i(\neq j)}T_{ij}.
}
This inequality yields the bound
\eq{
\tlT_{ii}-T_{ii}= \sum_{i(\neq j)}T_{ij}- \sum_{i(\neq j)}\tlT_{ij}\leq \frac{3\ep}{8} \sum_{i(\neq j)}T_{ij} \leq \frac{3\ep}{8}.
}
In conclusion, we obtain the bound from both above and below for any $i, j$:
\eq{
-\frac{7\ep}{16} \leq \tlT^{\bsy}_{ij}-T_{ij}\leq \frac{3\ep}{8},
}
where we explicitly manifest the $\bsy$-dependence of $\tlT$.

\bigskip

\figin{8cm}{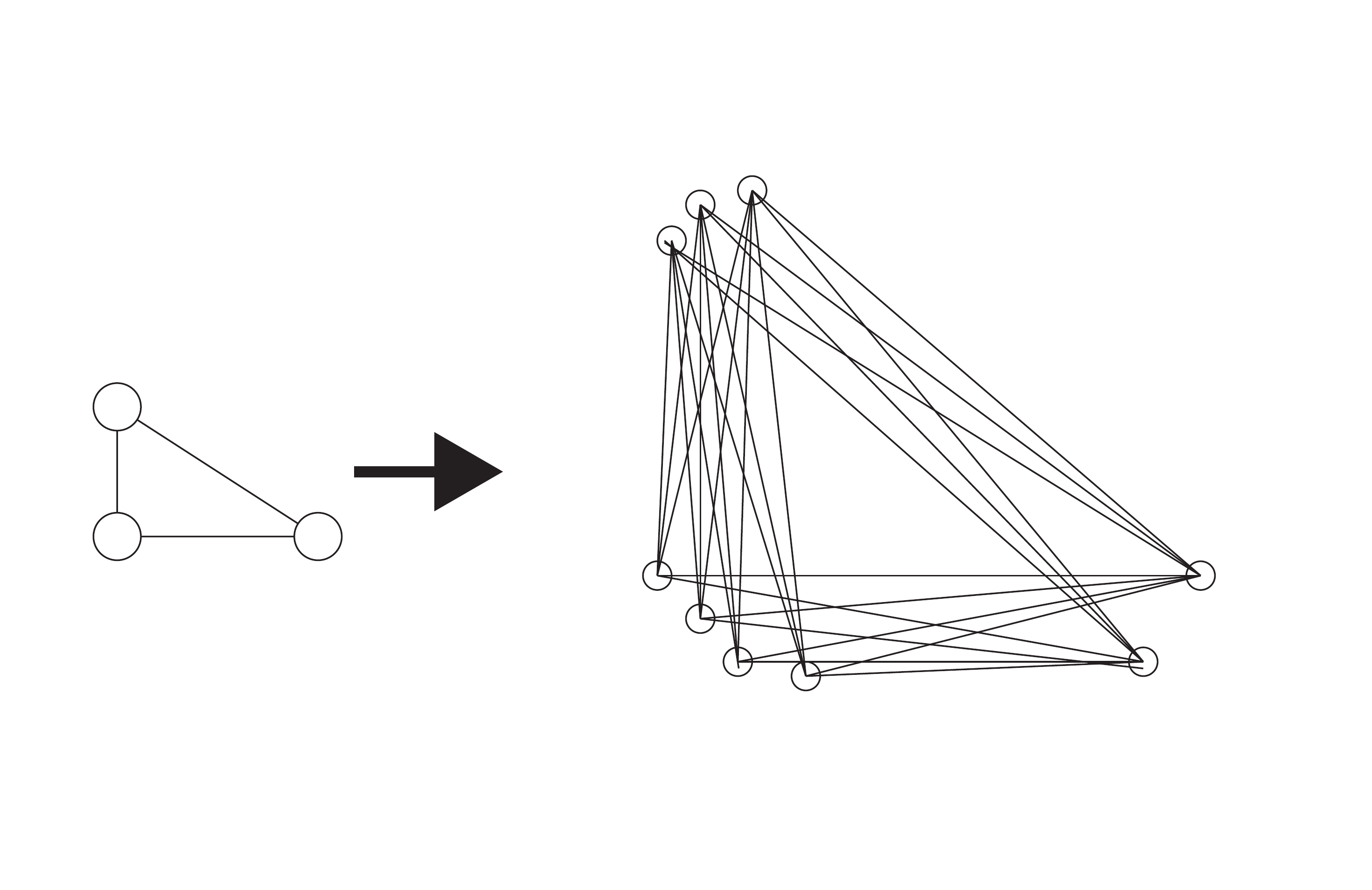}{
A schematic example of connections between islands (left) and transition paths between states in the islands (right).
A single line in the right figure implies a possible transition between the two states.
We remark that transitions between states in the same island are omitted.
}{TO-island}

We now explicitly write down the transition rate of TO.
Let $(i, (\bsx, m))$ be the initial state of the composite system, and define $\bsy:=\bsx-\bse_i$.
Then, TO can convert the initial state $(i, (\bsx^{i,\bsy}, m))$ to another state which can be expressed as $(j, (\bsx^{j,\bsy}, m'))$ with the same $\bsy$.
The state space of the composite system is decomposed into subsets of states with the same $\bsy$, and TO converts states inside the subset.
States inside a subset (fixed $\bsy$) are determined by the pair of $(i,m)$ ($i\in \{ 0,1,\ldots , d-1\}$ and $m\in \{ 0,1, \ldots, N(\bsx^{i, \bsy})\}$).

Since $\bsy$ is invariant under TO, we can construct doubly-stochastic matrices for each $\bsy$ independently.
Using $\tlT^{\bsy}$, we construct a doubly-stochastic matrix $D^{\bsy}$ for a fixed $\bsy$ as
\eqa{
D^{\bsy}_{(i,m)(j,n)}= \frac{\tlT^{\bsy}_{ij}}{N(\bsx^{i, \bsy})},
}{def-Dy}
which is independent of $m,n$
(See also \fref{TO-island}.
We, however, remark that this schematic does not draw the transition line for $i=j$ for visibility, though the definition of \eref{def-Dy} is valid for the case of $i=j$).
We demonstrate that the constructed matrix $D^{\bsy}$ is indeed a doubly-stochastic matrix:
\balign{
\sum_{i,m}D^{\bsy}_{(i,m)(j,n)}&=\sum_i N(\bsx^{i, \bsy}) \cdot \frac{\tlT^{\bsy}_{ij}}{N(\bsx^{i, \bsy})} =\sum_i \tlT^{\bsy}_{ij}=1, \\
\sum_{j,n}D^{\bsy}_{(i,m)(j,n)}&=  \frac{1}{N(\bsx^{i, \bsy})} \sum_j \tlT^{\bsy}_{ij} N(\bsx^{j, \bsy})=\frac{1}{\tlp^{\bsy}_i} \sum_j \tlT^{\bsy}_{ij} \tlp^{\bsy}_j \nt \\
 &=\frac{1}{\tlp^{\bsy}_i}\sum_j \tlT^{\bsy}_{ji} \tlp^{\bsy}_i =1.
}
The desired doubly-stochastic matrix $D$ is given by the sum of $D^{\bsy}$ over all $\bsy$:
\eq{
D:=\sum_{\bsy}D^{\bsy}.
}

\figin{10cm}{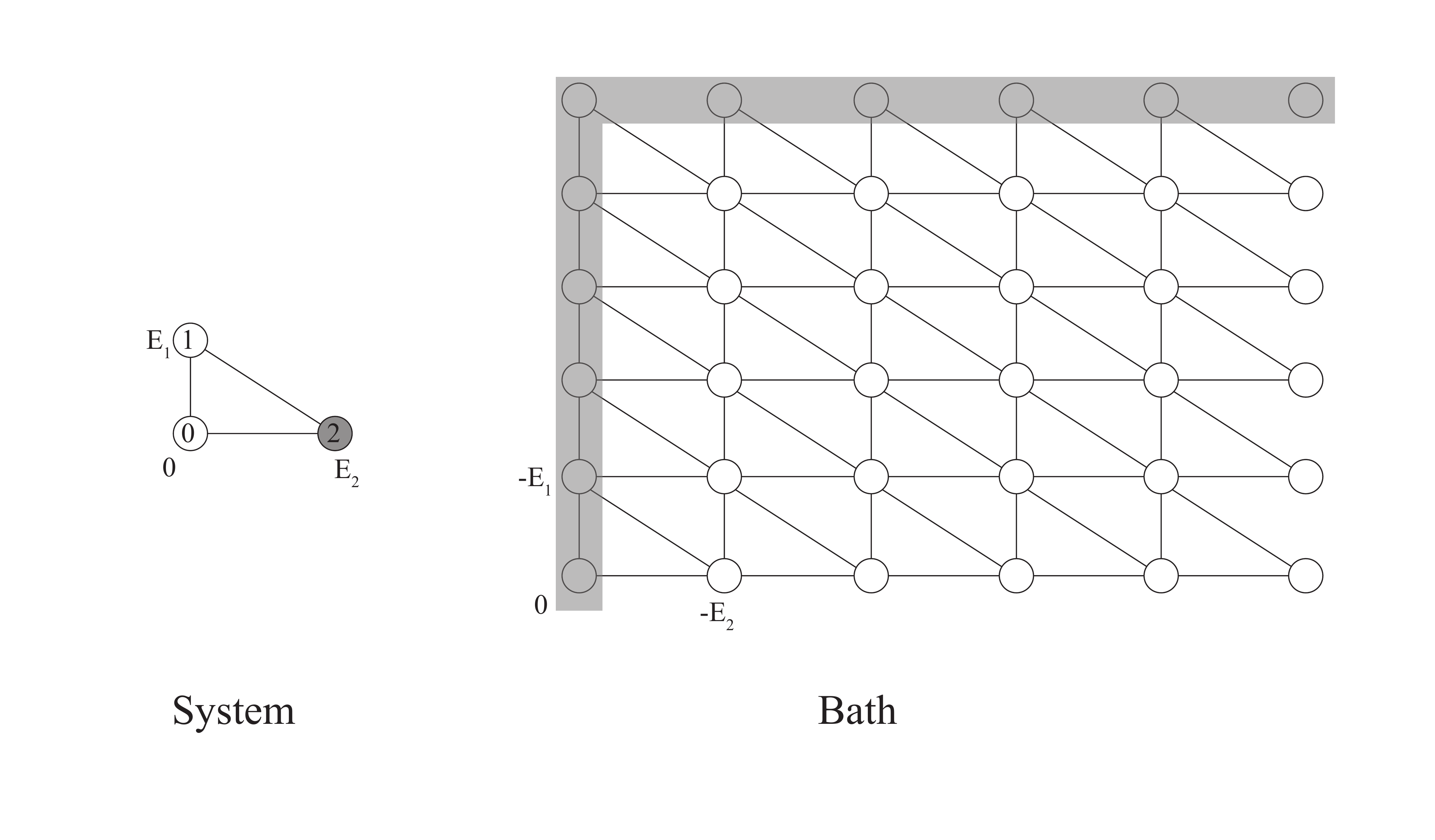}{
An example when the transition in the system cannot occur.
Suppose the initial state of the system is the state 2 (gray circle in the left figure).
If the initial island of the bath is one of the islands in the gray band, then the state cannot change.
We should evaluate the probability for these {\it bad} initial islands.
}{TO-edge}

We finally demonstrate that a TO given by $D$ indeed emulates the GPM $T$ with an accuracy $\ep$.
The dynamics contracted to the system is described by the following stochastic matrix $D'$:
\eq{
D'_{ij}=\sum_{\bsy}\sum_{m,n}D^{\bsy}_{(i,m)(j,n)}\frac{e^{\beta \bsE \cdot \bsx^{j,\bsy}}}{\sum_{\bsx'} N(\bsx')e^{\beta \bsE \cdot \bsx'} }=\sum_{\bsy}\tlT^{\bsy}_{ij}\frac{N(\bsx^{j,\bsy})e^{\beta \bsE \cdot \bsx^{j,\bsy}}}{\sum_{\bsx'} N(\bsx')e^{\beta \bsE \cdot \bsx'} }.
}
We evaluate $D'_{ij}$ from both above and below.
First, $D'_{ij}$ is bounded from above as
\eq{
D'_{ij}\leq \sum_{\bsy}\( T_{ij}+\frac{3\ep}{8}\) \frac{N(\bsx^{j,\bsy})e^{\beta \bsE \cdot \bsx^{j,\bsy}}}{\sum_{\bsx'} N(\bsx')e^{\beta \bsE \cdot \bsx'} }\leq T_{ij}+\frac{3\ep}{8}.
}
Next, we bound $D'_{ij}$ from below.
In this evaluation, we remark that $\bsx^{j,\bsy}$ takes only $L^{d-1}$ different points with fixed $j$, though the number of lattice points is $(L+1)^{d-1}$.
Owing to this fact, in case of some (unfortunate) initial islands the composite system cannot be converted to other states.
For example, suppose that the initial state of the system is in the gray circle of \fref{TO-edge}.
In this case, if the initial island is in the gray band of \fref{TO-edge}, then the state of the system cannot change.
We should estimate the probability of these initial states.
By taking this probability into account, Eqs.~\eref{ratio-approx} and \eref{cond-L} lead to
\eq{
\sum_{\bsy} \frac{N(\bsx^{j,\bsy})e^{\beta \bsE \cdot \bsx^{j,\bsy}}}{\sum_{\bsx'} N(\bsx')e^{\beta \bsE \cdot \bsx'} }\geq \frac{L^{d-1}\( 1-\frac{\ep}{32}\) }{(L+1)^{d-1} \( 1+\frac{\ep}{32}\)} \geq 1-\frac{\ep}{8},
}
which directly implies
\balign{
D'_{ij}&\geq \sum_{\bsy}\( T_{ij}-\frac{7\ep}{16}\) \frac{N(\bsx^{j,\bsy})e^{\beta \bsE \cdot \bsx^{j,\bsy}}}{\sum_{\bsx'} N(\bsx')e^{\beta \bsE \cdot \bsx'} } \nt \\
&\geq \( T_{ij}-\frac{7\ep}{16}\) \( 1-\frac{\ep}{8}\) \geq T_{ij}-\frac{9\ep}{16}>T_{ij}-\ep.
}
Hence, we conclude that $D'$ emulates $T$ with an accuracy $\ep$:
\eq{
\abs{D'_{ij}-T_{ij}}<\ep .
}

\section{Discussion}

In the first part, we presented a diagrammatic proof of the construction of the stochastic matrix $T$ satisfying $p'=Tp$ and $q'=Tq$ under the assumption $(p,q)\succ (p',q')$.
This proof is graphical and highly intuitive, which confirms the existence of the desired stochastic matrix $T$ with no doubt.
Corresponding analytical description in a mathematical form is also provided.
Our proof contains Birkhoff's theorem as its special case (both $q$ and $q'$ are the uniform distribution).
This simple constructive proof will be helpful in implementation of the stochastic matrix by numerical simulation.

In the second part, we presented an explicit construction of thermal operation which emulates any given Gibbs-preserving map.
In addition, we clarified the meaning of equivalence of the Gibbs-preserving map and thermal operation in the classical regime.
Although some Gibbs-preserving maps cannot be emulated by thermal operation exactly, any Gibbs-preserving map is emulatable by thermal operation with an arbitrary accuracy.
Our construction is based on the direction manifested in the paper of Horodecki and Oppenheim~\cite{HO}.
In fact, our proof can be understood as filling some vague or abstract points in the argument of Ref.~\cite{HO}.
Our construction accompanies evaluation of error (accuracy) for finite size systems, which tells the speed of convergence.
This findings will push further the recent recognition of the importance of finiteness of heat baths in quantum thermodynamics~\cite{TH, SOF, SM18, RAM}.

Simple constructive proofs help our simple and intuitive comprehension, which sometimes deepen our understanding, and therefore such alternative proofs are welcomed even if the problem has already been proven in some complicated ways.
A possible direction in this line is the problem of catalytic majorization (trumping).
In catalytic majorization, the state conversion with {\it catalyst} is investigated.
Catalytic majorization $p\prec_{\rm c} q$ is equivalent to $p\otimes r\prec q\otimes r$ with some proper catalyst state $r$.
It has been proven that the set of $\alpha$-Renyi entropies serves as monotones for catalytic majorization~\cite{Kli, Tur, AN}, while these proofs are highly complicated and clear understanding of catalytic majorization has still not yet been obtained.
Since catalytic majorization has various potential applications even restricted to quantum thermodynamic problems~\cite{Bra15, Mul}, simplified constructive proof of catalytic majorization will help our further understanding of majorization, which is a future problem.

\ack

The author is grateful to Takahiro Sagawa for his pedagogical lecture on majorization, which triggered this research.
The author also thanks Ryuji Takagi for careful reading and various helpful comments.
The author is supported by JSPS Grants-in-Aid for Scientific Research Grant Number JP19K14615.

\appendix

\renewcommand{\thesection}{Appendix.\Alph{section}}

\section{Some remarks on the results of Horodecki-Oppenheim}
 \appnum{\Alph{section}}
 
We here briefly review and comment on the argument of Horodecki and Oppenheim~\cite{HO}, which claims that the TO emulates any GPM with an arbitrary accuracy in (semi-)classical systems.
Although they treat semi-classical quantum systems, since it is equivalent to classical systems, in this section we describe their result in terms of classical probability distributions.

We first summarize their claim.
For any given $\delta>0$, they first suppose a very large heat bath and a set of states of the bath $K$ such that 
\eq{
\sum_{i\in K}\pG_i\geq 1-\delta
}
and satisfies the following four properties:
\be{
\renewcommand{\labelenumi}{(\roman{enumi})}
\item Almost all states in $K$ have energy in $[\bar{E}-\sqrt{\bar{E}}, \bar{E}+\sqrt{\bar{E}}]$ with some $\bar{E}$.
\item Let $N(E)$ be the number of states (degeneracy) with energy $E$. 
For any $i\in K$, the degeneracy of energy $E_i$ scales exponentially with respect to $E$:
\eq{
N(E_i)\geq e^{cE_i}
}
with a positive constant $c>0$.
\item Let $a, b$ be two states of the system with energy $E_a$ and $E_b$.
For any states of the system $a,b$ and any state of the bath $i\in K$, there exists a state of the bath $j\in K$ satisfying
\eq{
E_a+E_i=E_b+E_j.
}
\item For any state of the system $a$ and any state of the bath $i\in K$, the number of states of the bath with energy $E_i$ and $E_i-E_a$ obeys the Boltzmann's law approximately:
\eq{
\abs{\frac{N(E_i)e^{-\beta E_a}}{N(E_i-E_a)}-1}\leq \delta .
}
}
They {\it assume} the existence of such a bath, and using these conditions they construct a TO emulating a given GPM with an arbitrary accuracy.

Their main concern seems to be on the case of infinite baths which is obtained after taking the thermodynamic limit.
However, our interest is on a large but finite size heat bath and its asymptotic behavior (how TO converges to a given GPM as the size of the bath becomes larger).
Therefore, we here examine the validity of this assumption in case of a bath with large but finite system size.
The most problematic condition is (iii).
This assumption is justified by resorting the continuity of the energy spectrum of the bath.
However, any large but finite size heat bath does not have continuous energy spectrum.
This point becomes serious if the energy difference in the system has irrational ratio.

Before going to the irrationality problem, we first clarify the fact that (iii) in the literal meaning cannot be fulfilled in any finite size heat bath (regardless of irrationality).
The reason is very simple.
Since the system has finite number of states, $K$ is also a finite set.
On the other hand, (iii) requires that $K$ should have states $j$ with energy $E_j=E_i+n(E_a-E_b)$ for any $n\in \bbZ$, which is possible only when $K$ is an infinite set.
This is contradiction.
Hence, (iii) is violated at least at the edge of the energy band of $K$.

If all the energy of the system is rational numbers, then we can easily construct a heat bath where (iii) is violated only at the edge of the energy band of $K$.
In contrast, if the energy difference in the system has irrational ratio, the situation is much worse.
Consider a three state system $\{ a,b,c\}$ with $E_a=0$, $E_b=1$ and $E_c=\sqrt{2}$ as an example.
Let $\Emax_K$ and $\Emin_K$ be the maximum and minimum energy of states in $K$.
We decompose $K$ into $M=\lfloor (\Emax_K-\Emin_K)/(1+\sqrt{2})\rfloor$ sets $K_1, K_2,\ldots , K_M$, where states in $K_n$ ($n\neq M$) have energy in $[\Emin_K+ (n-1)(1+\sqrt{2}), \Emin_K+ n(1+\sqrt{2}))$ and states in $K_M$ have energy in $[\Emin_K+ (M-1)(1+\sqrt{2}), \Emax_K]$.
Let $E_i$ be an energy of a state in the set $K_n$ (i.e., $E_i\in [\Emin_K+ (n-1)(1+\sqrt{2}), \Emin_K+ n(1+\sqrt{2}))$).
We consider the following sequence of $e_m$:
\bi{
\item $e_0=E_i$.
\item If $e_m+1\in [\Emin_K+ (n-1)(1+\sqrt{2}), \Emin_K+ n(1+\sqrt{2}))$, then we set $e_{m+1}:=e_m+1$.
\item If $e_m+1\notin [\Emin_K+ (n-1)(1+\sqrt{2}), \Emin_K+ n(1+\sqrt{2}))$, then we set $e_{m+1}=e_m-\sqrt{2}$.
}
This sequence $\{ e_m\}$ has infinite length and stays in $[\Emin_K+ (n-1)(1+\sqrt{2}), \Emin_K+ n(1+\sqrt{2}))$.
In addition, all $e_m$ take different values for different $m$.
Hence, if (iii) is satisfied, for any $m$ the set $K_n$ should contain a state with energy $e_m$.
However, this is impossible for a finite set $K_n$, which implies the violation of (iii) in all $K_n$ ($n=1,2,\ldots , M$).
In other words, in case of irrational energy difference ratio, (iii) is violated in the bulk of the energy band of $K$.

\bigskip

We shall compare the assumed requirements in Ref.~\cite{HO} and our construction of thermal operation.
The condition \eref{cond-L} along with the structure of the state space of the bath is a counterpart of the requirement (iii).
The structure of the state space ensures that most of the states (islands) satisfy the condition (iii), and the inequality \eref{cond-L} bounds the number of states not satisfying the condition (iii).
The condition \eref{ratio-approx} corresponds to the requirement (iv).
We do not put conditions corresponding to the requirements (i) and (ii) explicitly.

\end{document}